\UseRawInputEncoding
\documentclass[11pt]{article}

\usepackage[top=1in,bottom=1in,left=1in,right=1in]{geometry}

\usepackage{graphicx}
\usepackage{amsmath}
\usepackage[colorlinks,urlcolor=blue,citecolor=blue]{hyperref}

\title{\textbf{A Novel Tool for the Accurate and Affordable Early Diagnosis of Pancreatic Cancer via Machine Learning and Bioinformatics}}

\author{Siya Goel \\
\texttt{siyagoel24@gmail.com} \\
Precollege Research Opportunities, Purdue University
\and
Clark Gedney \\
\texttt{cgedney@purdue.edu} \\
Biological Sciences, Purdue University
\and
Jean Honorio \\
\texttt{jhonorio@purdue.edu} \\
Computer Science, Purdue University}

\date{\vspace{-0.3in}}

\begin{document}

\maketitle

\begin{abstract}

Pancreatic cancer (PC) is the fourth leading cause of cancer death in the United States due to its five-year survival rate of 10\%.
Late diagnosis, affiliated with the asymptomatic nature in early stages and the location of the cancer with respect to the pancreas, makes current widely-accepted screening methods unavailable.
Prior studies have achieved low (70-75\%) diagnostic accuracy, possibly because 80\% of PC cases are associated with diabetes, leading to misdiagnosis.
To address the problems of frequent late diagnosis and misdiagnosis, we developed an accessible, accurate and affordable diagnostic tool for PC, by analyzing the expression of nineteen genes in PC and diabetes.
First, machine learning algorithms were trained on four groups of subjects, depending on the occurrence of PC and Diabetes.
The models were analyzed with 400 PC subjects at varying stages to ensure validity.
Naive Bayes, Neural Network and K-Nearest Neighbors models achieved the highest testing accuracy of around 92.6\%.
Second, the biological implication of the nineteen genes was investigated using bioinformatics tools.
It was found that these genes were significantly involved in regulating the cytoplasm, cytoskeleton and nuclear receptor activity in the pancreas, specifically in acinar and ductal cells.
Our novel tool is the first in the literature that achieves a PC diagnostic accuracy of above 90\%, having the potential to significantly improve the detection of PC in the background of diabetes and increase the five-year survival rate.

\end{abstract}

\section{Introduction}

Pancreatic
cancer (PC) is a disease in which malignant acinar and ductal cells are formed in the pancreatic tissue~\cite{li_pancreatic_2004}.
The pancreas is a gland located behind the stomach and in front of the spine~\cite{bardeesy_pancreatic_2002}.
The pancreas has two key functions: (i) helping with digestion (exocrine) and (ii) regulating blood sugar (endocrine)~\cite{kim_intercellular_2001}.
More specifically, in the exocrine function, the pancreatic duct secretes enzymes which help in breaking down fats, carbohydrates and proteins~\cite{zoppi_exocrine_1972,ballian_islet_2007}.
The endocrine function of the pancreas consists of producing insulin which lowers blood glucose levels and glucagon which raises blood glucose levels~\cite{molina_pancreas_2020}.
Maintaining this blood sugar level is important in the functioning of key organs such as the brain, liver and kidneys~\cite{chan_loss_2016}.

\begin{figure}
\begin{center}
\includegraphics[width=0.5\linewidth]{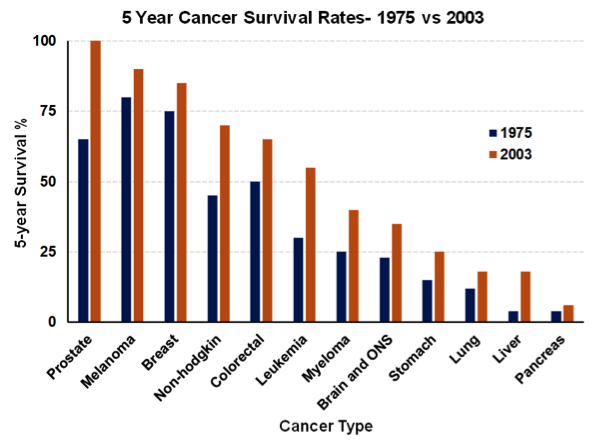}
\vspace{-0.05in}
\caption{The five-year diagnosis rate of different cancers~\cite{carpelan-holmstrom_does_2005}.}
\label{fig:fiveyearsurvival}
\end{center}
\end{figure}

Even though only 56,770 Americans are diagnosed with PC (3\% of all cancers), 47,740 people die from it, thus making PC the fourth leading cause of cancer death, causing 9\% of deaths in cancer patients~\cite{hidalgo_pancreatic_2010}.
The main reason for this high mortality rate is the difficulty of PC diagnosis~\cite{nakazawa_difficulty_2007}.
The majority of people suffering from PC are diagnosed at stage IV, i.e., the stage when the disease has metastasized and there are signs of symptoms~\cite{porta_exocrine_2005}.
In fact, compared with many other cancers, the combined five-year survival rate for PC (i.e., the percentage of all patients who are living five years after diagnosis) is relatively low at just 10\%, compared to the survival rate of lung cancer as well as breast cancer (Figure~\ref{fig:fiveyearsurvival})~\cite{quaresma_40-year_2015}.

PC is difficult to diagnose early because the pancreas is located deep inside the abdomen~\cite{hidalgo_pancreatic_2010}.
Further, patients usually do not have any symptoms until the cancer has reached later stages or has already spread to other organs.
As a result, early-stage tumors cannot be seen or felt by health care providers during routine physical exams~\cite{martin_irreversible_2013}.
In addition, there are no specific, cost-effective screening tests that can easily and reliably find early-stage PC in people who are asymptomatic~\cite{zhang_challenges_2018}.
Patients usually do not have any symptoms until the cancer has reached later stages or has already spread to other organs.
Studies suggest that if the cancer were detected at an early stage (Stage I or II) when surgical removal of the tumor was possible, the five-year survival rate would be 40\%~\cite{kamisawa_pancreatic_2016}.
However, only 15-20\% of people are diagnosed at an early stage~\cite{zhang_challenges_2018}.
According to prior studies, up to 80\% of PC patients have diabetes which increases the risk of PC by 8-fold~\cite{pezzilli_is_2013}, which suggests that diabetes could be a precursor of PC~\cite{andersen_pancreatitis-diabetes-pancreatic_2013}.
However, because of inaccuracies of the current diagnostic methods, 20\% of PC cases are mistakenly diagnosed as diabetes~\cite{sah_new_2013}.
Therefore, in order to aid in the detection of PC tumors at early stages, we developed a diagnostic tool for PC by distinguishing between gene expression in PC and diabetes.
A potential solution leading to a more accurate diagnosis could be to discover differentially and similarly expressed genes between diabetes, PC, and normal cells~\cite{bardeesy_pancreatic_2002}.

\subsection{Related Work}

To the best of our knowledge, our work is the first machine learning study to focus on early diagnosis and to compare PC and diabetes for gene signatures.
That is, none of the prior works that we discuss in this section pertain to early diagnosis.

Image diagnosis is one of the two main methods used to diagnose PC~\cite{clarke_preoperative_2003}.
Unfortunately, image diagnosis is considered ineffective, due to image blur, resolution and the appearance of tumors and visual change in later stages, making early diagnosis difficult~\cite{chari_detecting_2007}.
This makes PC imaging inaccurate, expensive and unattainable~\cite{mcmahon_pancreatic_2001}.

A study done by~\cite{cai_pancreas_2016} examined 82 images of pancreatic segmentation for diagnosis.
The images obtained were axial CT slices of the pancreas, which were analyzed by a variety of convolutional neural networks.
However, these algorithms only achieved 70\% accuracy in diagnosis~\cite{cai_pancreas_2016}.
Another study done by~\cite{zhou_fixed-point_2017} focused on pancreatic segmentation through pictures obtained from MRI scans from 78 subjects.
Two convolutional neural networks were made to distinguish the difference between normal and cancerous tissue, achieving a 73.2\% accuracy~\cite{zhou_fixed-point_2017}.
A third study done by~\cite{dmitriev_classification_2017} took a different approach and classified the four most common cyst types.
The model consisted of two parts (i) a probabilistic random forest classifier, which analyzes manually selected quantitative features and (ii) a convolutional neural network trained to discover high-level imaging features for a better differentiation.
The data used contained 134 abdominal CT scans, achieving an accuracy of 83.6\%~\cite{dmitriev_classification_2017}.

Besides imaging methods, another method used to diagnose PC is through genetic expression analysis~\cite{alldinger_gene_2005}.
According to a recent survey conducted by~\cite{yan_importance_2019}, only three studies have used machine learning for gene expression in PC.
All of these studies consisted of a limited amount of samples (i.e., 175).
The accuracy of the three machine learning studies were 77\%, 80\% and 83\%~\cite{yan_importance_2019}.
It is important to note that the above studies were not accounting for the misdiagnosis of PC with diabetes, as gene expression between the two groups are similar.
In addition, in all of the above studies, 2,000 parameters or genes were used to predict the binary output label (i.e., whether the subject has PC or not) which might be affected by overfitting.
We argue that in order to avoid overfitting, only a subset of genes need to be considered in order to provide an accurate diagnosis~\cite{ko_mirna_2018}.

\section{Materials and Methods}

Our aim is to develop an early diagnostic tool of PC in an accessible, accurate, timely and affordable manner by analyzing expression of few (tens of) genes in PC and diabetes.
More specifically, our goal is to:
\begin{itemize}
\item obtain an accuracy of over 90\% for the diagnosis for PC in both early and late stage PC
\item understand the difference between the four groups of subjects depending on the occurrence of PC and diabetes
\item provide a simple and accessible way to diagnose PC in order to make it more available to the public by creating a simple yet accurate algorithm
\item provide a cheap and affordable diagnosis
\end{itemize}

Our design consists of the two phases described below:

\paragraph{Phase 1.}

400 samples (i.e., subjects) from open-source, annotated datasets were used in order to ensure the validity of our study over different gene expression levels.
A series of machine learning classifiers (i.e., Logistic Regression, K-Nearest Neighbors, Random Forest, Naive Bayes, Neural Network) were applied on four groups of subjects, depending on the occurrence of PC and diabetes.
Nineteen genes were chosen from independent studies.

\paragraph{Phase 2.}

We found the biological implication between the nineteen genes analyzed and significant genetic function in PC such as cytoplasmic structure, formation of actin cytoskeleton and nuclear receptor activity.

In what follows, we describe the above two phases in detail.

\subsection{Phase 1: Applying Machine Learning Algorithms}

Five different classifiers: Logistic Regression, K-Nearest Neighbors (KNN), Random Forest, Naive Bayes and Neural Network, were applied to observe whether the accuracy of diagnosis of early PC could be improved.
The dataset retrieved for this study is comprised of four groups of subjects:
\begin{itemize}
\item S11: subjects with Diabetes and PC
\item S10: subjects with Diabetes, without PC
\item S01: subjects without Diabetes, with PC
\item S00: subjects without Diabetes, without PC
\end{itemize}

\begin{table}
\begin{center}
\begin{small}
\caption{Summary of data retrieved for this study.}
\label{tab:data}
\vspace{0.1in}
\begin{tabular}{p{3in}p{2.25in}}
\hline
\textbf{Dataset} & \textbf{Number of Samples} \\
\hline
The Cancer Genome Atlas: project PAAD & S11: 63 samples \newline
 S01: 91 samples \\
\hline
Gene Expression Omnibus: datasets GSE22309 & S00: 72 samples \newline
 S10: 32 samples \\
\hline
Gene Expression Omnibus: datasets GSE15932, GSE16515, GSE14245 and GSE49515 & S00: 34 samples \newline
 S10: 22 samples \newline
 S01: 51 samples \newline
 S11: 33 samples \\
\hline
\textbf{Total Count} & 398 samples \newline
 S00: 106 samples \newline
 S10: 54 samples \newline
 S01: 142 samples (early stage PC: 47, late stage PC: 95) \newline
 S11: 96 samples (early stage PC: 34, late stage PC: 62) \\
\hline
\end{tabular}
\end{small}
\end{center}
\end{table}

In the above dataset, one subject corresponds to one sample or observation.
Thus, in this manuscript, we interchangeably refer to subjects or samples.
A total of 398 samples were retrieved and consisted of 106 samples from S00, 54 samples from S10, 142 samples from S01, and 96 samples from S11.
The sources of data included The Cancer Genome Atlas (project PAAD) and the Gene Expression Omnibus (datasets GSE22309, GSE15932, GSE16515, GSE14245 and GSE49515) (Table~\ref{tab:data}).
The Cancer Genome Atlas samples were filtered to be the ones that (i) mentioned whether or not the patient had diabetes and (ii) mentioned the stage of PC.
The Gene Expression Omnibus samples were chosen based on the following keywords: ``pancreatic cancer'' and ``pancreatic adenocarcinoma''.
The ``organism'' filter was set to ``Homo sapiens'' and the ``study type'' filter was set to ``expression profiling by array''.
The datasets were chosen to be the ones that (i) mentioned whether or not the patient had diabetes, and (ii) mentioned the stage of PC.

Nineteen genes were found to be significantly overexpressed in PC through literature and were selected as features for the five classifiers.

\begin{table}
\begin{center}
\begin{small}
\caption{Nineteen genes selected as features from the literature.}
\label{tab:genes}
\vspace{0.1in}
\begin{tabular}{lll}
\hline
\textbf{Gene Name} & \textbf{Gene ID} & \textbf{Reference} \\
\hline
ABHD12 & ENSG00000100997 & \cite{kind_structural_2019} \\
ABHD14B & ENSG00000114779 & \cite{carr_differentiation_2012} \\
ABHD2 & ENSG00000140526 & \cite{yu_retinoic_2019} \\
ACTA2 & ENSG00000107796 & \cite{li_pancreatic_2019} \\
ACTB & ENSG00000075624 & \cite{guo_actb_2013} \\
ACTN1 & ENSG00000072110 & \cite{rajamani_identification_2016} \\
ACTN4 & ENSG00000130402 & \cite{honda_biological_2015} \\
ACTR1A & ENSG00000138107 & \cite{tang_identification_2019} \\
ACTR2 & ENSG00000138071 & \cite{rauhala_silencing_2013} \\
ADAR & ENSG00000160710 & \cite{aravindan_polyphenols_2015} \\
ADPRHL2 & ENSG00000116863 & \cite{zhang_discovery_2017} \\
ADRA2A & ENSG00000150594 & \cite{liu_reg3a_2015} \\
ADRM1 & ENSG00000130706 & \cite{long_development_2012} \\
ALDH1A1 & ENSG00000165092 & \cite{kim_reversing_2013} \\
ALDH9A1 & ENSG00000143149 & \cite{kim_reversing_2013} \\
ALKBH5 & ENSG00000091542 & \cite{fujii_alkbh2_2013} \\
ALKBH7 & ENSG00000125652 & \cite{fujii_alkbh2_2013} \\
ANXA11 & ENSG00000122359 & \cite{shen_aspirin:_2013} \\
APOC1 & ENSG00000130208 & \cite{zelcer_liver_2006} \\
\hline
\end{tabular}
\end{small}
\end{center}
\end{table}

Through literature, it was found that nineteen genes play a significant role in monitoring actin production and acinar and ductal cell function, and tend to be overexpressed in PC.
Thus, they were selected as features for the five classifiers.
The nineteen genes are shown in Table~\ref{tab:genes}.

We used the Python programming language with libraries such as NumPy, Panda and SciKit-Learn.
The data from sets S00 and S01 were named ``No Diabetes'' and the data from sets S10 and S11 were named ``Diabetes''.
Both datasets ``Diabetes'' and ``No Diabetes'' contain input features (i.e., the expression of the nineteen genes) and a binary output label (i.e., whether the subject has PC or not).

We tested five different classifiers and considered different hyperparameters to be tuned for each of them.
For the Logistic Regression classifier, we considered the hyperparameter C (inverse of regularization strength) to be either of ten values between 0.01 and 1000 in a geometric scale.
For the Random Forest classifier, we considered the hyperparameter n\_estimators (number of trees) to be either 1, 2, 3, 5, 8, 10, 15, 20, 25 or 30.
For the K-Nearest Neighbors, we considered the hyperparameter n\_neighbors (number of neighbors) to be either 10, 20, 30 or 40.
For the Naive Bayes classifier, we considered the hyperparameter var\_smoothing (variance added for calculation stability) to be either of ten values between 10-7 to 10 in a geometric scale.

In what follows, we describe the procedure that was applied independently for each classifier.
For cross-validation, the ``No Diabetes'' dataset was randomly split into an 80\% training dataset and a remaining 20\% validation dataset.
We trained the classifier for different values of the hyperparameter on the training set.
We then computed the accuracy on the validation set.
We kept track of the value of the hyperparameter that performed best, i.e., the value that led to the highest accuracy in the validation set.
This process was then iterated 50 times, and we chose the value of the hyperparameter that performed best most of the 50 times.
The classifier with the best hyperparameter value chosen above was trained on the entire ``No Diabetes'' dataset.
We then computed the following performance metrics on our test dataset, i.e., the ``Diabetes'' dataset:
\begin{itemize}
\item $\displaystyle{ \text{Accuracy} = \frac{\text{TP} + \text{TN}}{\text{TP} + \text{FP} + \text{TN} + \text{FN}} }$
\item $\displaystyle{ \text{Precision} = \frac{\text{TP}}{\text{TP} + \text{FP}} }$
\item $\displaystyle{ \text{Recall} = \frac{\text{TP}}{\text{TP} + \text{FN}} }$
\item $\displaystyle{ \text{F2 Score} = \frac{5 \times \text{Precision} \times \text{Recall}}{4 \times \text{Precision} + \text{Recall}} }$
\item AUC is the area under the receiver-operator-characteristics curve, i.e., the area under the curve defined by Recall and $\displaystyle{ \text{Specificity} = \frac{\text{TN}}{\text{TN} + \text{FP}} }$.
\end{itemize}
In the above formulas, TP is the number of ``true positives'', i.e., subjects with PC, predicted as such.
TN is the number of ``true negatives'', i.e., subjects without PC, predicted as such.
FP is the number of ``false positives'', i.e., subjects without PC, predicted as having PC.
FN is the number of ``false negatives'', i.e., subjects with PC, predicted as not having PC.

The entire process described above was repeated 100 times, which allowed us to assess the mean and standard error of each of the performance metrics.

\subsection{Phase 2: Bioinformatics Study}

\begin{figure}
\begin{center}
\includegraphics[width=0.4\linewidth]{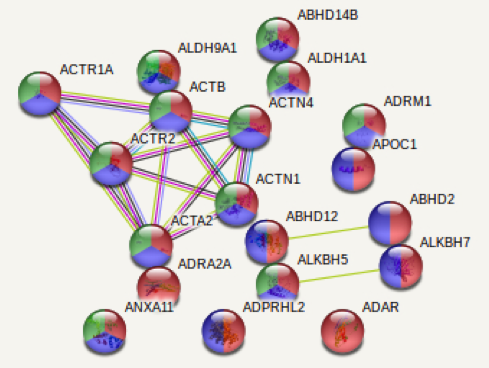}
\vspace{-0.05in}
\caption{Involvement of the genes in acinar and ductal cells in the cytoplasmic structure (red), actin cytoskeleton formation (blue) and nuclear receptor activity (green).}
\label{fig:involvementgenes}
\end{center}
\end{figure}

\begin{figure}
\begin{center}
\includegraphics[width=0.5\linewidth]{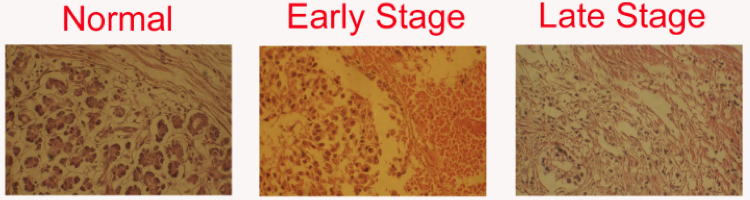}
\vspace{-0.05in}
\caption{Representative 10x images from different stages of pancreatic cancer in our dataset.}
\label{fig:stages}
\end{center}
\end{figure}

The study protocol has been approved by the Internal Review Board of Purdue University, and informed consent was obtained from participants.
To test the biological implication behind our findings, the nineteen genes were put into String-DB~\footnote{\texttt{https://string-db.org/}} (Figure~\ref{fig:involvementgenes}).
The connections between the genes were analyzed including the significance of the gene in the morphology and functionality of the nucleus, cytoplasm and cell membrane, as well as expression in the pancreas, specifically in acinar and ductal cells.
This was done using the GeneCards~\footnote{\texttt{https://www.genecards.org/}} database which contains data from Entrez Gene, UniProtKB, PathCards and BioGPS.

To understand the physiological difference of the key components the nineteen genes were involved in, we used a PA804a tissue microarray from US Biomax.
A Nikon E600 microscope was then connected to a Nikon D500 camera.
Pictures of cells of each group in the microarray were taken, and acinar and ductal cells were identified~\cite{longnecker_2014}.
From an initial set of 300 images, we excluded the ones with less than 10\% occurrence of acinar and ductal cells, using the ImageJ~\footnote{\texttt{https://imagej.nih.gov/ij/}} software.
The remaining thirty images consisted of 5 tissue samples without PC but with diabetes, 5 samples of without PC neither diabetes), 5 samples of early stage PC with diabetes, 5 samples of early stage PC without diabetes, 5 samples of late stage PC with diabetes, and 5 samples of early stage PC without diabetes.
We compared the cell area and aspect ratio (major axis/minor axis) of the above thirty images (Figure~\ref{fig:stages}).

\section{Results}

\begin{figure}
\begin{center}
\includegraphics[width=0.5\linewidth]{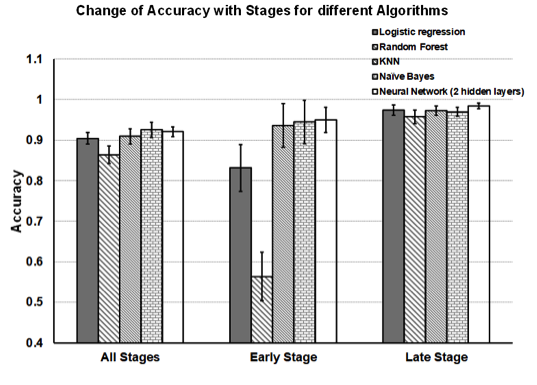}\includegraphics[width=0.5\linewidth]{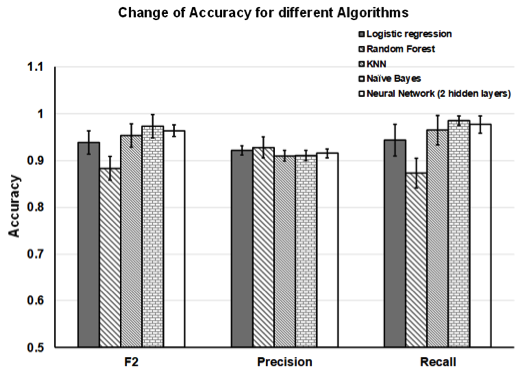} \\
\includegraphics[width=0.5\linewidth]{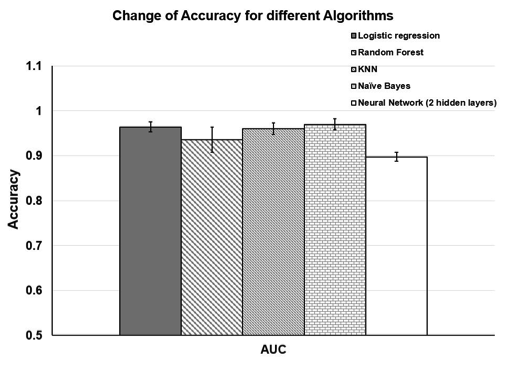}
\vspace{-0.1in}
\caption{Performance metrics for the different classifiers.}
\label{fig:performance}
\end{center}
\end{figure}

\begin{figure}
\begin{center}
\includegraphics[width=0.5\linewidth]{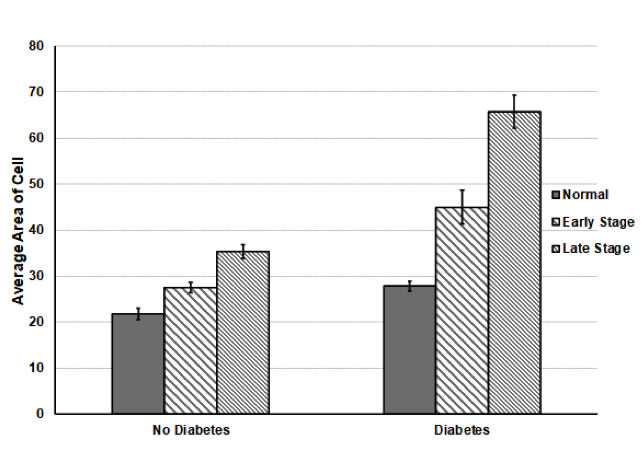}\includegraphics[width=0.5\linewidth]{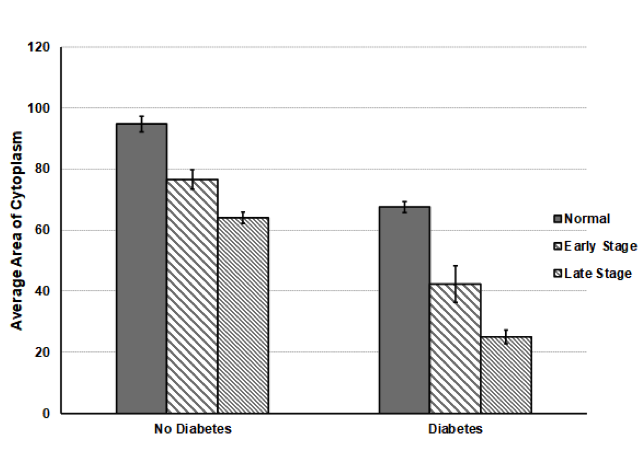} \\
\includegraphics[width=0.5\linewidth]{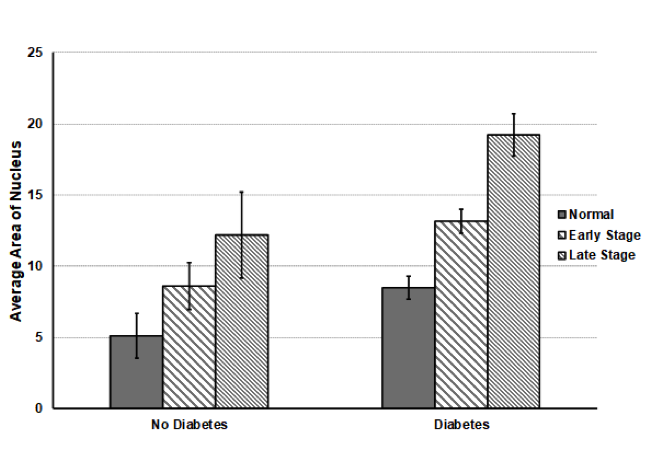}
\vspace{-0.05in}
\caption{Area of the cell, cytoplasm and nucleus for samples with and without diabetes and for different stages of PC.
(Each image contained 80-100 acinar or ductal cells.)}
\label{fig:area}
\end{center}
\end{figure}

\begin{figure}
\begin{center}
\includegraphics[width=0.5\linewidth]{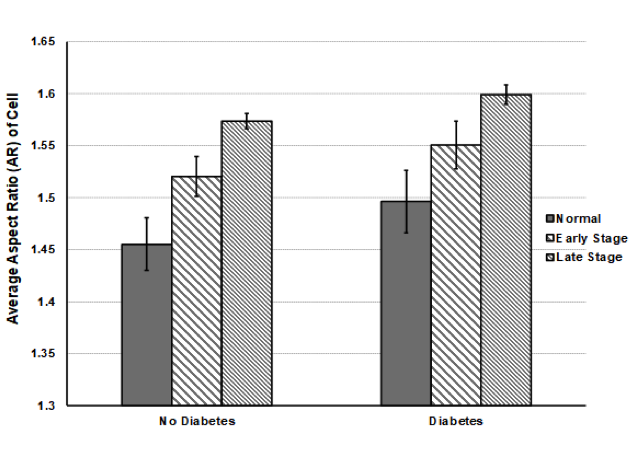}\includegraphics[width=0.5\linewidth]{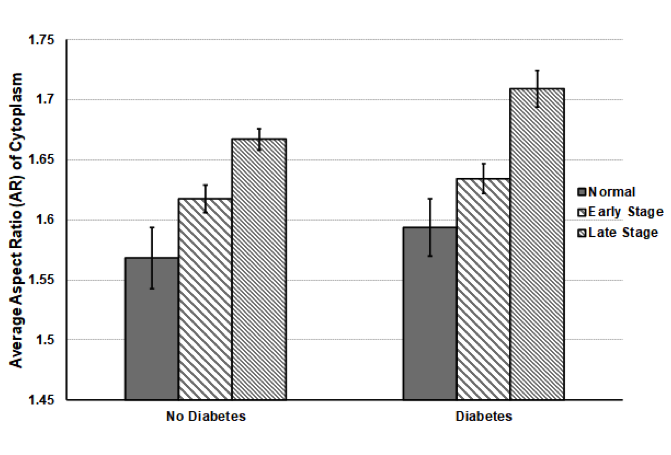} \\
\includegraphics[width=0.5\linewidth]{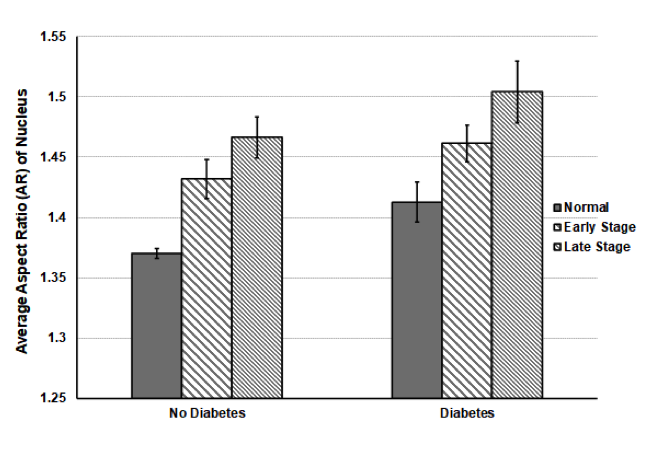}
\vspace{-0.05in}
\caption{Aspect ratio of the cell, cytoplasm and nucleus for samples with and without diabetes and for different stages of PC.
(Each image contained 80-100 acinar or ductal cells.)}
\label{fig:aspectratio}
\end{center}
\end{figure}

\begin{figure}
\begin{center}
\includegraphics[width=0.5\linewidth]{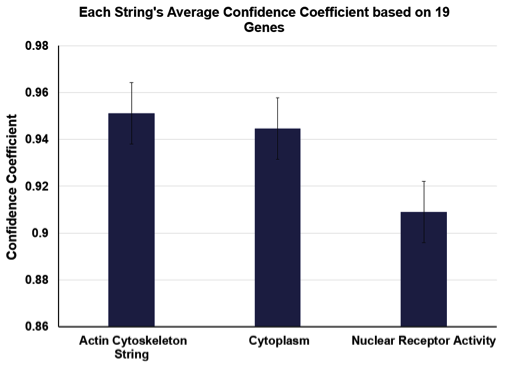}\includegraphics[width=0.5\linewidth]{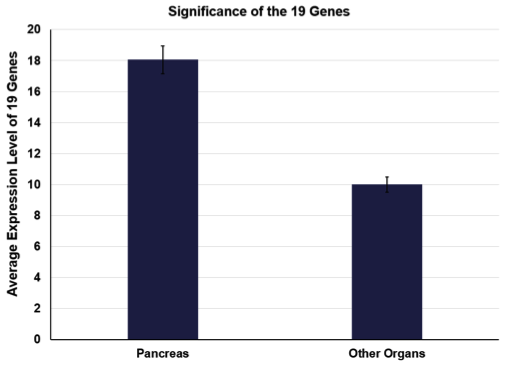}
\vspace{-0.1in}
\caption{Genes involvement in the strings' activity on the GeneCards database.
The results suggest that the difference in expression of the nineteen genes may cause the changes shown in Figures~\ref{fig:area} and~\ref{fig:aspectratio}.}
\label{fig:proteinsequences}
\end{center}
\end{figure}

Overall accuracy, accuracy in late stage PC subjects, accuracy in early stage PC subjects, precision, recall, AUC and F2 scores of the KNN, Naive Bayes and Neural Network models were all $>$0.90, proving to be significantly accurate, according to acceptable diagnostic standards~\cite{richardson_beyond_2006}.
Early stage accuracy for Logistic Regression and Random Forest models is $<$0.90, and thus the methods are inaccurate for early stage accuracy.
However, accuracy in late stage PC subjects is $>$0.90, showing high accuracy for late diagnosis (Figure~\ref{fig:performance}).

As PC progresses, there is an increase in the average area of the cell and nucleus, a decrease in the area of the cytoplasm, and an increase in the aspect ratio in all three structures (Figures~\ref{fig:area} and~\ref{fig:aspectratio}).
The standard error bars (mean $\pm$ 0.05) in the area of the cell, cytoplasm, and nucleus with diabetes did not overlap, showing that the increase or decrease was statistically significant.
The standard error bars in the aspect ratio of the cell, cytoplasm, and nucleus with diabetes did not overlap, again, showing that the increase in aspect ratio was statistically significant.
However, the area and aspect ratio without diabetes overlapped, showing that the increase in area and aspect ratio was not statistically significant.

The nineteen genes are involved in the cytoskeleton, cytoplasm and nucleus of the acinar and ductal cells at high confidence intervals (Figure~\ref{fig:proteinsequences}).
Following~\cite{morey_fallacy_2016}, since the confidence coefficients were all $>$0.90, the relationship between the nineteen genes and strings' activity are significant
The average expression of the 19 genes is significantly higher in the pancreas compared to the average expression of these genes in other organs.
This is because the standard error bars between the two groups do not overlap in Figure~\ref{fig:proteinsequences}.

\section{Discussion}

All classifiers except random forest achieved the goal of obtaining an accuracy above 90\% to distinguish between the four groups of subjects, proving that the algorithms were optimal in diagnosing PC.
All the classifiers achieved an accuracy greater than 90\% for late stage diagnosis; however, the logistic regression and random forest did not do so for early diagnosis.
Therefore, the best classifiers were the KNN, Naive Bayes and Neural Network due to their high performance in both late and early diagnosis.
The logistic regression and random forest classifiers behaved poorly possibly due to the simplicity of logistic regression and high model complexity of random forests.

One of the main reasons why these classifiers performed well is because the nineteen genes were identifiers of both early diagnosis and misdiagnosis.
The aforementioned genes may control the area in the cytoplasm, the nucleus, and the acinar or ductal cell which change based on the presence of PC and/or diabetes.
The nucleus grows as cancer progresses because there is an increase in the need for ribosomes and polysomes which have proteins necessary for the cell growth process, a chain upregulated in cancer~\cite{huang_ductal_2015}.
The proteins thus cause the cell to grow rapidly and the cytoplasm decreases due to nuclear enlargement~\cite{son_glutamine_2013}.
This in turn causes the aspect ratio to get larger, as cell growth generates irregularity in shape~\cite{gaviraghi_pancreatic_2011}.
The same changes occur in type 2 diabetes as the cell and nucleus grow to produce leptin, a hormone that maintains fat content~\cite{meek_leptin_2013}.
However, this leptin production is less rapid compared to the uncontrollable cell growth in cancer, causing less drastic changes in cellular, nuclear, and cytoplasmic areas and aspect ratio~\cite{zhao_lysine-5_2013}.
These differences suggest (i) that the cellular, nuclear, and cytoplasmic changes started early on in the pancreas and (ii) that differences in these areas and aspect ratios are different depending on the occurrence of PC and/or diabetes.

Literature supports the involvement of these nineteen genes with PC and diabetes as PC samples have been affiliated with higher expression levels in the ACT genes, the main gene groups considered for our machine learning algorithms.
The ACT genes are found to be upregulated in 80\% of PC subjects and are significantly upregulated in diabetes~\cite{honda_biological_2015}.
This is because these genes change the shape of the cytoskeleton due to their involvement with alpha actins~\cite{riviere_novo_2012}.
These actins are a diverse group of cytoskeletal proteins, including the alpha and beta spectrins and dystrophins, which are involved in binding the actin to the membrane~\cite{perrin_actin_2010}.
Thus, our analysis suggests that the ACT genes are linked to changes in the cellular shape of both PC and diabetes, in acinar and ductal cells.

Our novel tool is the first in the literature that achieves a PC diagnostic accuracy of above 90\%.
Together with the belief that the five-year survival rate would be 40\% if the cancer were detected at an early stage~\cite{kamisawa_pancreatic_2016}, we conclude that our tool can potentially increase the five-year survival rate to 36\% (i.e., 40\% $\times$ 90\%).

\section{Concluding Remarks}

The additional applications of the project include public use, affordability and continuous improvement.
Our results allow for many people to use our diagnostic tool due to affordable and easy techniques to extract gene expression, like RT-qPC.
This diagnostic tool is also more affordable compared to the current CT and MRI scans used to diagnose PC.
These algorithms will improve further if applied to the real world as more data will be available, making better conclusions.
The bioinformatics analysis conducted provides a better understanding of the connection between the nineteen genes and their functionality.
This genetic understanding could give rise to precision medicine in order to control the expression of the nineteen genes, leading to specialized PC treatment.

Currently, the classifiers cannot tell the difference between early and late diagnosis.
To help improve practicality and provide patients for an accurate treatment, the classifiers can be trained to predict not just whether the patient has PC or not, but also the stage of PC.
Further, to solidify the significance of these genes in PC, specific genes out of the nineteen can be analyzed using in situ hybridization.
Since String-DB shows that all these genes are directly correlated, we conjecture that differences in fluorescence in a couple of genes (e.g., ACTA2 and ABHD12) will prove that these genes are differentially expressed in the four different groups.

\paragraph{Acknowledgements.}

The authors would like to thank Prof. Brittany Allen-Petersen for the very helpful feedback, and the PreCollege Research Opportunities Program at Purdue University.

\bibliographystyle{abbrv}
\bibliography{references}

\end{document}